\documentclass[aps,twocolumn,showpacs,preprintnumbers,prl]{revtex4}
\usepackage{graphicx}
\usepackage{epsfig,latexsym}

\newcommand{\beq}{\begin{equation}}
\newcommand{\eeq}{\end{equation}}
\newcommand{\be}{\begin{equation}}
\newcommand{\ee}{\end{equation}}
\newcommand{\ber}{\begin{eqnarray}}
\newcommand{\eer}{\end{eqnarray}}
\newcommand{\berr}{\begin{eqnarray*}}
\newcommand{\eerr}{\end{eqnarray*}}
\newcommand{\ba}{\begin{array}}
\newcommand{\ea}{\end{array}}

\newcommand{\jpsi}{J / \psi}

\newcommand{\old}[1]{}

\begin{document}
\title{Color Screening Melts Quarkonium}

\author{\'Agnes M\'ocsy \email{mocsy@bnl.gov}} \affiliation{ RIKEN-BNL
  Research Center, Brookhaven National Laboratory, Upton NY 11973, USA }
\author{P\'eter Petreczky\email{petreczk@bnl.gov}} \affiliation{
Physics Department, Brookhaven National Laboratory, Upton NY 11973, USA }

\begin{abstract}
We calculate quarkonium spectral functions in a quark-gluon plasma
using a potential model based on full QCD lattice calculations of the
free energy of static quark-antiquark pair. We estimate the binding energy and the thermal width of different quarkonium states. 
The estimated upper limit for the dissociation temperatures is considerably lower than the ones suggested in the recent literature. 
\end{abstract}
\preprint{BNL-NT  07/26 }
\preprint{RBRC-683}
\maketitle

One of the most important features of the quark gluon plasma is the screening of static chromo-electric fields \cite{mclerran}. It has been argued that screening above the 
deconfinement temperature $T_c$ is strong enough to lead to the dissolution of the $J/\psi$ state, which can then signal quark-gluon plasma formation in heavy ion collisions \cite{MS86}. The fact that the  $J/\psi$-suppression pattern observed  in SPS and RHIC experiments is still not understood \cite{exp}, 
serves as a motivation for this work.  
In particular, our  aim is to relate quarkonium suppression to  deconfinement and color screening.

Because of the large quark mass $m = m_{c,b} \gg \Lambda_{QCD}$, the
velocity $v$ of heavy quarks in the bound state is small, and the
binding effects in quarkonia at zero temperature can be understood in
terms of  non-relativistic potential models \cite{Lucha91}. 
More recently, the potential has been derived from QCD using a sequence of
effective field theories: Non-relativistic QCD (NRQCD), an effective
theory where all modes above the scale $m$ are integrated out, and
potential NRQCD, an effective theory in which
all modes above the scale $mv $ are integrated out \cite{nrqcd_rev}.

Based on the success of the potential model at zero temperature, and on 
the idea that color screening implies modification of the inter-quark
forces, potential models have been used to try to understand quarkonium properties at finite T  \cite{MS86,potmod}. 
To discuss properties and dissolution of quarkonium states at
finite temperature spectral functions have to be considered.
Using lattice QCD, charmonium correlators have been
calculated, and the corresponding spectral functions have been
extracted using the Maximum Entropy Method (MEM)
\cite{umeda02,asakawa04,datta04,swan,jako06}.
The MEM at zero temperature can reconstruct the basic features of the
spectral functions: the ground state, the excited states, and the
continuum \cite{jako06}.  At finite temperature, however, the extraction of the
spectral functions becomes difficult, because the length of the Euclidean time is
limited by the inverse temperature. 

In a recent paper \cite{mocsy07} we calculated  the
spectral function using a potential model for the non-relativistic Green's function. This detailed analysis shows that spectral functions
calculated in a potential model combined with perturbative QCD can describe the available lattice data on quarkonium correlators
at zero and at finite temperature in QCD with no light quarks \cite{mocsy07}. Charmonium states were found to dissolve 
at temperatures significantly lower than quoted in lattice QCD studies. In the present work we extend our analysis to the case of real QCD
with one strange quark and two light quarks. Furthermore, using lattice results on color screening 
we derive upper bounds for the dissociation temperatures of various quarkonium states in the quark-gluon plasma which are considerably lower than previous estimates. 

In the energy domain of the resonances and continuum threshold the spectral functions are calculated from the non-relativistic Green's functions as \cite{mocsy07}
\begin{eqnarray}
&
\displaystyle
\sigma(\omega)=\frac{2 N_c}{\pi} {\rm Im} G^{nr}(\vec{r},\vec{r'},E)|_{\vec{r}=\vec{r'}=0}\, ,\\[2mm]
&
\displaystyle
\sigma(\omega)=\frac{2 N_c}{\pi}\frac{1}{m^2} {\rm Im} \vec{\nabla} \cdot \vec{\nabla'} G^{nr}(\vec{r},\vec{r'},E)|_{\vec{r}=\vec{r'}=0}\, ,
\label{green_sc} 
\end{eqnarray}
for $S$-wave, and $P$-wave quarkonia, respectively. Here $E=\omega-2 m~$, and the number of colors is
$N_c=3$. The non-relativistic Green's function satisfies the Schr\"odinger equation
$
\left [ -\frac{1}{m} \vec{\nabla}^2+V(r)-E \right ] G^{nr}(\vec{r},\vec{r'},E) = \delta^3(r-r')\, .
$
The numerical method for solving this equation is presented in \cite{mocsy07}.  
At large energies, away from the threshold, the non-relativistic treatment is clearly not applicable. 
The spectral function in this domain, however,  can be calculated using perturbation theory \cite{mocsy07}. 

\begin{figure}
\includegraphics[width=8cm]{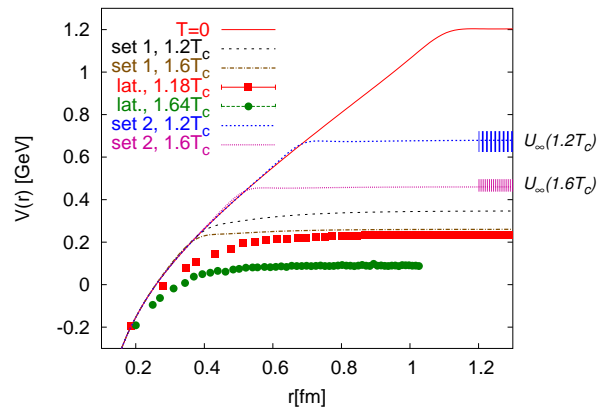}
\caption{The two choices of the potential (denoted as set 1 and set 2) used in our analysis for $1.2T_c$ and $1.6T_c$. 
We also show the free energy of static quark anti-quark pair 
 calculated in 2+1 flavor QCD on the lattice \cite{kostya}.}
\label{fig:pot}
\end{figure}

To calculate $G^{nr}(\vec{r},\vec{r'},E)$ we need  to specify the potential $V(r)$ in the Schr\"odinger equation. 
A Coulomb plus linear form, phenomenologically successful in describing quarkonia spectrum at $T=0$, also gives a good parametrization 
of the lattice data on the static quark-antiquark potential for  $0.1~\mbox{fm} < r < 1~$fm.  Only at distances $r<0.1~$fm, not relevant for quarkonia  
studies, the effect of the running coupling is important \cite{okacz}. At large distances, $r_{med} \simeq 1.1~$fm, the linear growth of the potential stops due to string breaking \cite{bali05}. 
At high temperatures we expect that the effective range of the potential will be reduced and the interactions are exponentially screened at large distances. At sufficiently short distances, on the other hand, the interaction between the heavy quark and antiquark 
is temperature independent \cite{okacz}. Motivated by the above, we use the following parametrization for the potential
\begin{equation}
V(r,T)= \left\{ \begin{array}{ll}
                   -\frac{\alpha}{r} + \sigma r, & r<r_{med}(T)\, , \\[3mm] 
                   V_\infty(T)-\frac{\alpha'(T)}{r} e^{-\mu(T) r}, & r > r_{med}(T) \, .
                  \end{array}
          \right.
\label{pot}
\end{equation}
Following the recent 2+1 flavor lattice QCD analysis \cite{kostya}, we fix $\alpha=0.385$ and $\sigma=1.263/r_0^2~$, with $r_0=0.469~$fm 
being the Sommer-scale determined in \cite{gray05}. At zero temperature we choose $r_{med}=1.1~$fm and screening mass of $\mu=0.4~$GeV. 
The later is motivated by the fact that the heavy-light meson pair is in the isosinglet channel. This choice then corresponds to $V_\infty \simeq 1.2~$GeV. 
The potential is shown in Fig.~\ref{fig:pot}. 
For calculations of the charmonium spectral functions we have included in the potential also a relativistic spin-independent correction, estimated
to be $-0.8\sigma/(m^2r)$ \cite{bali97}. 
The zero temperature potential with the above choice of the parameters gives a fairly good description of the quarkonium spectrum.

At finite temperature we fix the parameters in (\ref{pot}) utilizing the information available from lattice QCD on the free energy of static heavy quark-antiquark pair.  
Free energy calculations are done in pure gluodynamics, 3-flavor and 2-flavor  QCD \cite{okacz,petrov04,okacz05}, and preliminary results are also available in 
the physically relevant case of one heavy strange quark 
and two light quarks  \cite{kostya} (quark masses correspond to pion mass of about $200$~MeV). 
These calculations show that the free energy 
is temperature independent for distances $r<0.4~\mbox{fm}/(T/T_c)~$, while for distances $r T>0.8$ i
t is exponentially screened, with a screening mass estimated to be  $\mu(T)=1.4(1) \sqrt{1+N_f/6} g(T) T$ \cite{okacz,okacz05}. 
Here $g(T)$ is the 2-loop $\overline{MS}$ gauge coupling at scale $2 \pi T$. 
It was found also that at infinite separation of the heavy quark and antiquark the free energy approaches a constant $F_{\infty}(T)~$. 
At high temperatures perturbation theory
tells us that $F_{\infty}(T)$ behaves like $-\alpha_s^{3/2} T$ \cite{me_hard} (lattice calculations confirm this behavior \cite{okacz,petrov04,okacz05,kostya}). 
For $T>T_c$  we assume that $F_{\infty}(T)=a/T-bT$, where $a$ and $b$ are temperature-independent constants. The first term can be viewed as the contribution 
from a non-perturbative dimension 2 gluon-condensate \cite{megias}, while the second term is the perturbative entropy contribution. 
This parametrization successfully describes the lattice data on $F_{\infty}(T)$ in pure gauge theory \cite{megias}, and we checked that it works
very well  for $T>1.1T_c$ also in 2+1 flavor QCD.  In our numerical analysis we use the value $T_c=0.192~$GeV determined in \cite{Cheng:2006qk}.
Although the 2+1 flavor lattice data on the free energy of static quark-antiquark used here were obtained on coarse lattices with temporal extent $N_{\tau}=4$, discretization errors in this quantity are small \cite{petrov04}.

It is important to note, that since it contains an entropy contribution, the free energy itself is not the potential. It can, nevertheless, provide some constraints 
on the parameters of (\ref{pot}). In particular, the discussion above implies 
$0.4~\mbox{fm}/(T/T_c)<r_{med}<0.8/T~$, and due to the negative entropy contribution $V_\infty(T)>F_{\infty}(T)$.  
We use $V_\infty=a/T$ and $\mu= 1.4\sqrt{1+N_f/6} g(T) T~$.
The values of $\alpha'(T)$ are determined in \cite{okacz05}. The above described choice for  $V_\infty, \alpha'$ and $\mu$, 
together with a requirement for the smoothness of the potential, predetermines the value of $r_{med}$. This  
turns out to be close to $0.4~\mbox{fm}/(T/T_c)$, the value determined in lattice simulations. This physically motivated potential, labeled as set 1, is shown 
for different temperatures in Fig~\ref{fig:pot}. For comparison we display also the lattice data on the free energies.   
Denoted as set 2 in Fig~\ref{fig:pot} an alternative choice, an upper bound on
the finite temperature potential is shown: When fixing $r_{med}=0.8/T$ the smoothness of the potential 
determines the value of $V_{\infty}(T)$. It turns out that this value is close to $U_{\infty}(T)$, the internal energy of the isolated static quark. We know that $U_{\infty}(T)$ 
provides an upper bound on the possible value of $V_{\infty}(T)~$, because it contains
all possible interactions of heavy quark with the medium. The values of  $\mu(T)$ and $\alpha'(T)$ were fixed as in set 1.
The numerical results on the spectral functions are mostly determined by the value of $V_{\infty}(T)$ and are insensitive to other details of the potential (see discussion in \cite{mocsy07} also on the stability of the numerical solution).

\begin{figure}[h]
\begin{minipage}[h]{7cm}
\epsfig{file=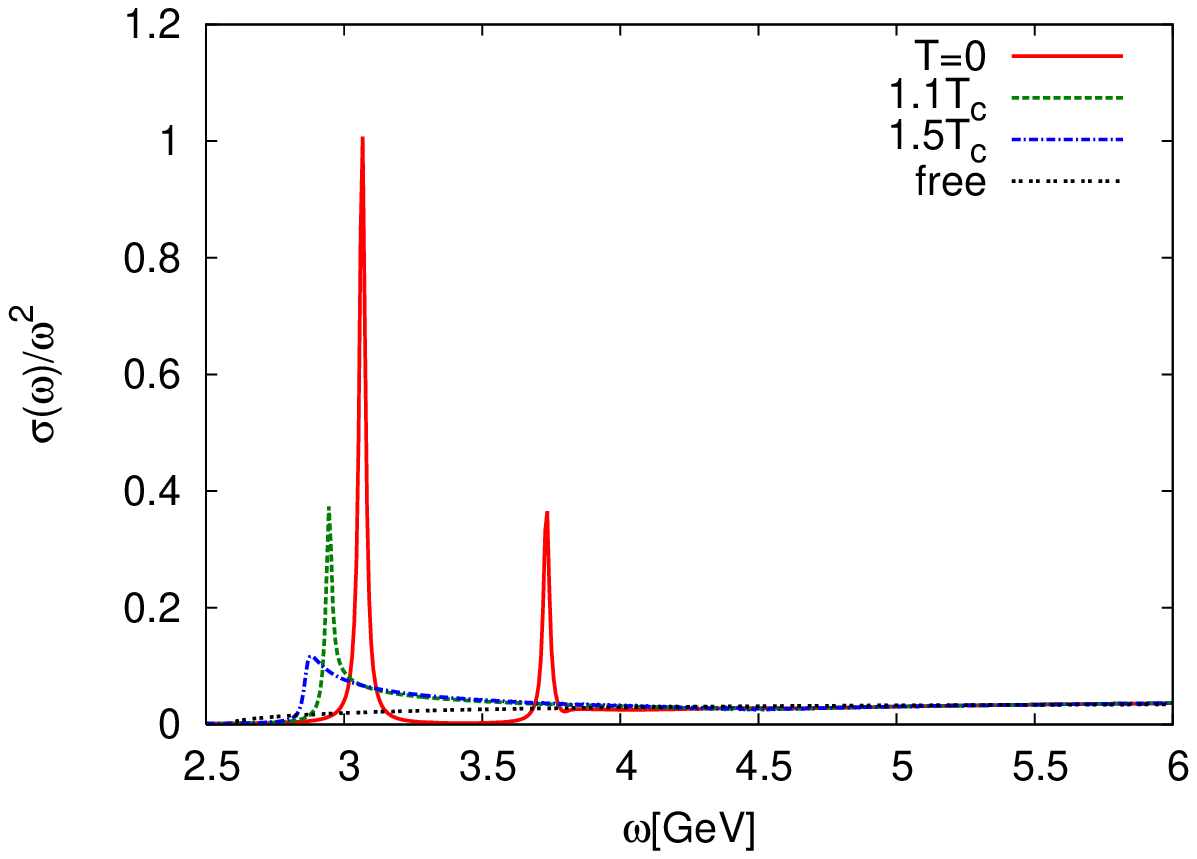,height=45mm}
\end{minipage}\\
\begin{minipage}[h]{7cm}
\epsfig{file=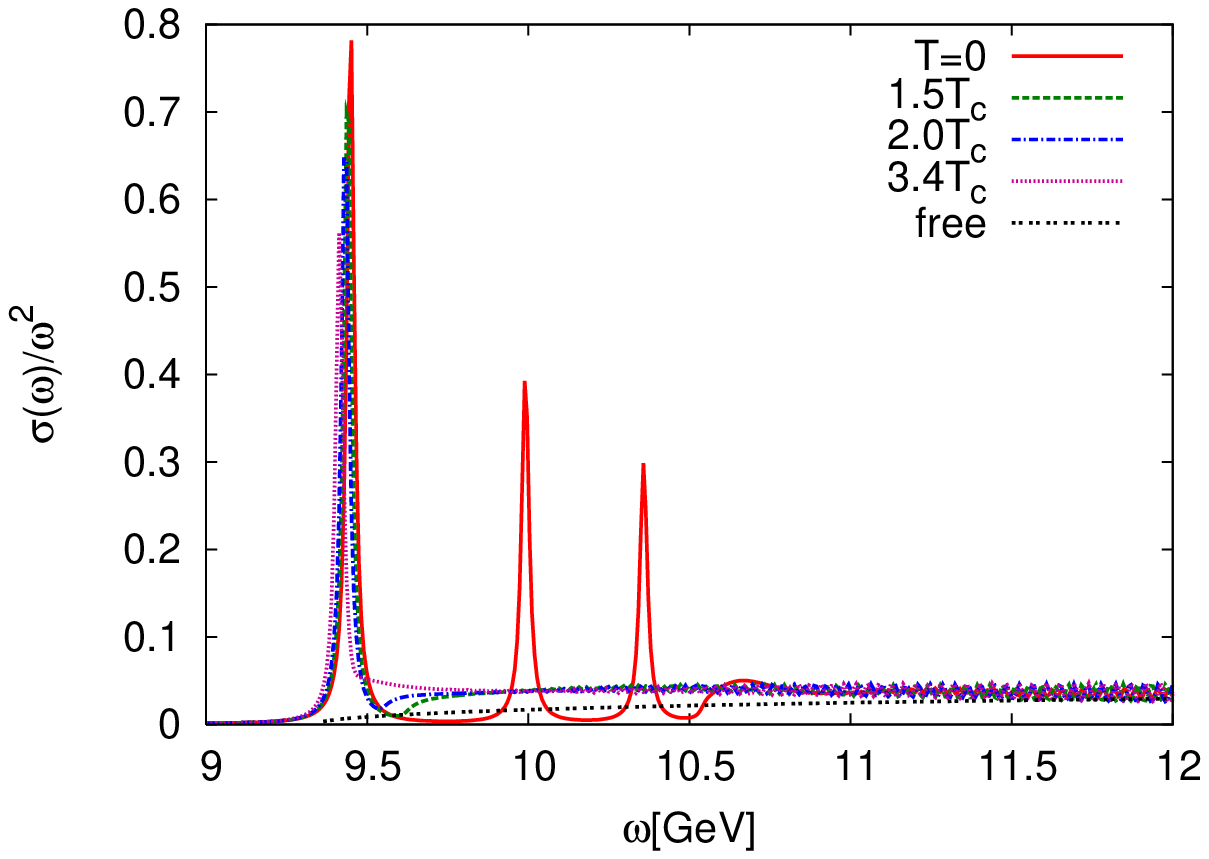,height=45mm}
\end{minipage}
\caption{S-wave charmonium (upper panel) and bottomonium (lower panel) spectral functions at different temperatures.}
\label{fig:spf}
\end{figure}

In what follows we present numerical results for these two choices of the potential, and derive an upper bound
for  the dissociation temperatures. 
In Fig.~\ref{fig:spf} we show the spectral functions above the deconfinement temperature. 
The S-wave charmonium spectral function at $T=1.1T_c$ exhibits a resonance peak with very small binding energy 
of about $0.014~$GeV.
Here, and in what follows, we define the binding energy as the distance between the peak position 
and the continuum threshold, $E_{bin}=2 m_{c,b}+V_{\infty}-M~$, $M$ being the
resonance mass. Above $1.1T_c$ the charmonium spectral functions show no resonance-like structures, meaning that all
charmonium states are dissolved.  Even though there are no resonances, the spectral function is strongly enhanced compared 
to the non-interacting case. This is  also illustrated in Fig.~\ref{fig:spf}.  
We would like to point out, that this threshold enhancement compensates for the 
dissociation of the states, and thus dramatic changes seen in the spectral function are not reflected in the correlation function (for a detailed 
discussion see \cite{mocsy07}). Also, the strong enhancement in the threshold region is an indication that the heavy quark and antiquark remain correlated.  
In the case of bottomonium we see only the ground state above deconfinement, all other states are dissolved. Furthermore, as Fig.~\ref{fig:spf}  shows,  there 
is no significant change in the peak position of the ground state up to $2T_c$.  
Even though seemingly the resonance structure persists to temperatures  even higher than this, the binding energy of the state is significantly reduced.  

When the binding energy of a resonance drops below the temperature the state is weakly bound, and thermal 
fluctuations can destroy it by transfering energy and exciting the quark anti-quark pair into the continuum. The rate of this excitation, or equivalently the width
of the quarkonium states, is determined by the binding energy \cite{dima95}.
Therefore,  in order to provide an upper bound on the dissociation temperature we need to estimate an upper
bound for the binding energy. To do this, we calculate quarkonium spectral functions for the set 2 potential, providing the maximum
possible binding still consistent with the lattice results on screening.  We find that with this choice of the potential the S-wave charmonium spectral function has 
resonance-like structures up to $\sim 1.6T_c$. Furthermore, we also see resonance like structures in the bottomonium spectral functions corresponding to the $1P$ and $2S$
states. In the upper panel of Fig. \ref{fig:binding} we show the corresponding binding energies of the different quarkonium states. 
Let us note, that in the past quarkonium widths at finite temperature have been calculated 
using perturbative QCD and the Boltzmann-approximation, assuming an ideal quark-gluon plasma. See \cite{wong07} for a recent analysis. 
For quarkonium sizes realized in nature the validity of the
perturbative calculations of the quarkonium-gluon cross section is doubtful. Furthermore, the Boltzmann-approximation breaks down if the binding energy is smaller
than the temperature. In \cite{dima95} the quarkonium dissociation rate due to thermal activation into the continuum has been estimated non-perturbatively, 
using a resonance plus a continuum model for the spectral function. The thermal dissociation rate $\Gamma(T)$ has a particularly simple form in two limits \cite{dima95}: The limit of
large and small binding, respectively: 
\begin{figure}
\includegraphics[width=6.5cm]{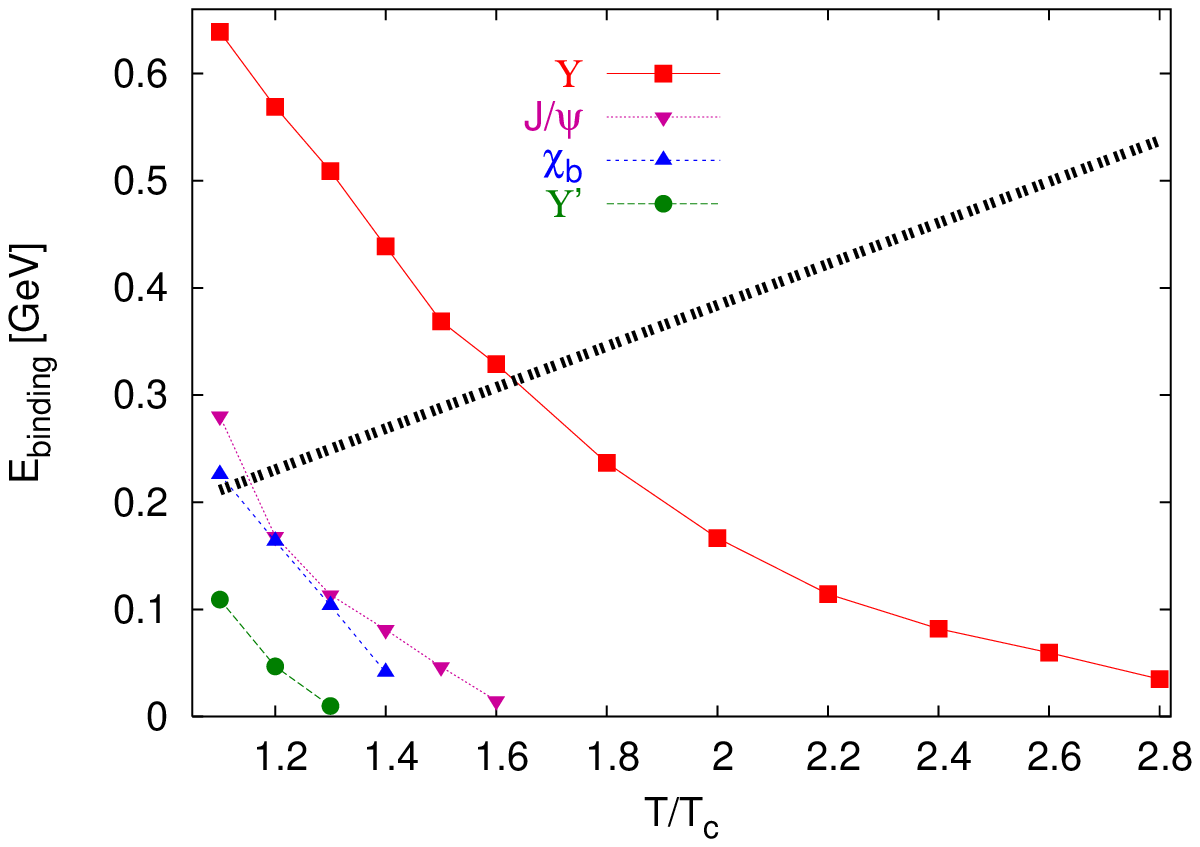}
\includegraphics[width=6.5cm]{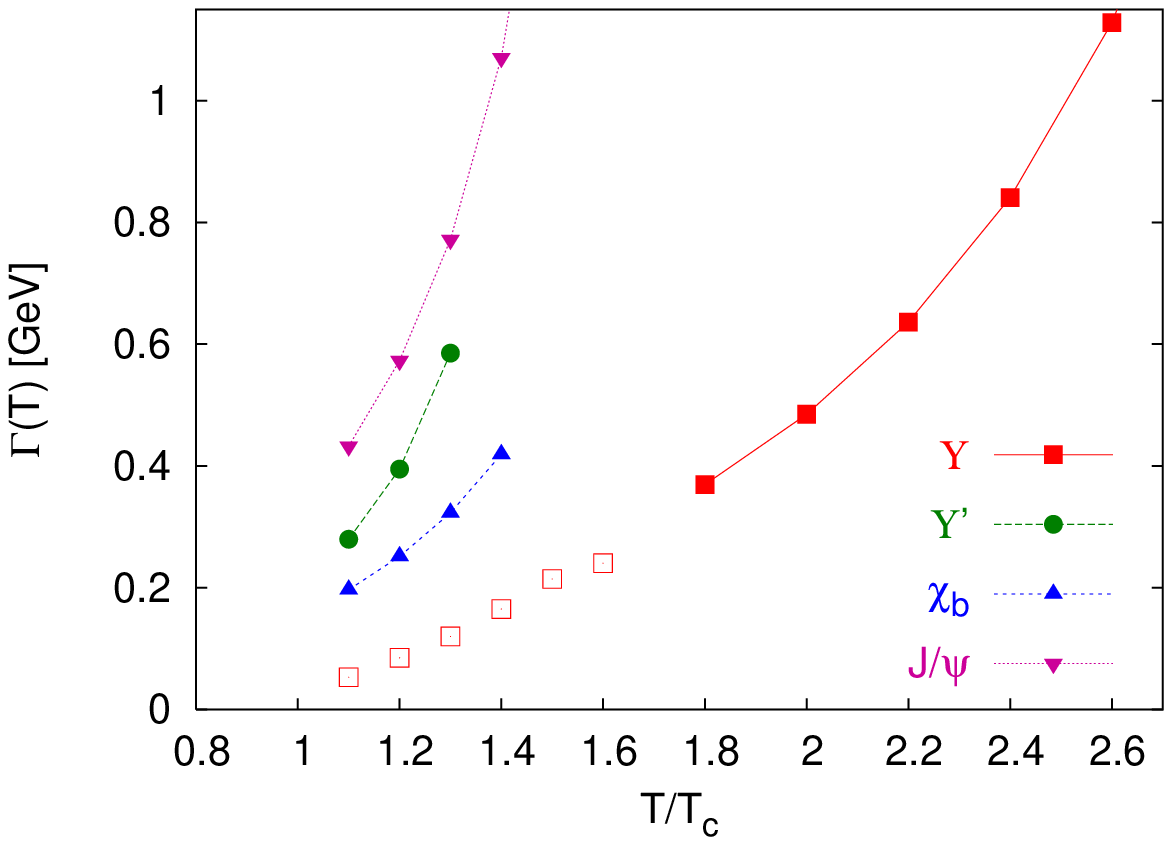}
\caption{An upper limit of the binding energy of different quarkonium states (top) and the quarkonium width
(bottom).  The open squares show the width of the 1S bottomonium state multiplied by six for better visibility,
which has been calculated in the limit of small binding.}
\label{fig:binding}
\end{figure}
\begin{eqnarray}
&
\Gamma(T)=\frac{(LT)^2}{3\pi}Me^{-E_{bin}/T}\, ,~E_{bin}\gg T \, ,
\label{large}\\
&
\Gamma(T)=\frac{4}{L}\sqrt{\frac{T}{2\pi M}}\, ,~ E_{bin}\ll T \, .
\label{small}
\end{eqnarray}
Here $L$ is the size of the spatial region of the potential, given by the distance from the average 
quarkonium radius to the top of the potential, i.e. $L=r_{med}-\langle r^2\rangle^{1/2}$. 
From the top panel of Fig. \ref{fig:binding} it is clear that for $T>1.1T_c$ all quarkonium states have binding energy smaller
than the temperature, with the exception of the $\Upsilon(1S)$ state. Their width can thus be estimated 
using (\ref{small}). 
The results are shown in the lower panel of Fig. \ref{fig:binding}. Note that all of these states have widths larger than $200~$MeV, and are therefore
likely to be dissociated in the plasma.
In the case of the $\Upsilon(1S)$  for $T<1.6T_c$ we use (\ref{large}) to estimate the width, which is found to be smaller than $40$MeV.  This is in fairly good agreement with the 
perturbative estimate of \cite{wong07}. For temperatures $T>1.6T_c$ even the $1S$ bottomonium 
is in the regime of small binding, and its width becomes very large by $2.6T_c$ (see Fig. \ref{fig:binding}). 
Note that uncertainty in the value of $U_{\infty}$ in the lattice calculations, indicated as a band in Fig.  \ref{fig:pot}, leads to uncertainty in the binding energy estimate of about $10\%$. 
When the thermal width is significantly larger than the binding energy the resonance 
structure seen in our calculation will not be observable
in reality. We define the dissociation temperature as the smallest temperature where no resonance structure can be seen in the spectral function. 
The upper limit for the dissociation temperatures of the 
quarkonium states we determine by posing the conservative quantitative condition $\Gamma(T) \ge 2 E_{bin}(T)$. The 
corresponding dissociation temperatures are summarized in Table \ref{tab:diss}.
A less conservative criterion $\Gamma(T) \ge E_{bin}(T)$ would reduce the dissociation temperature by roughly $10 \%$.

\begin{table}[hftb]
\renewcommand{\arraystretch}{0.81}
\begin{center}
\begin{minipage}{8.5cm} \tabcolsep 5pt
\begin{ruledtabular} 
\begin{tabular}{ccccccc}
 state&$\chi_c$&$\psi'$&$\jpsi$&$\Upsilon'$&$\chi_b$&$\Upsilon$\\ \hline 
 $T_{dis}$&$\le T_c$& $\le T_c$ &$1.2T_c$&$1.2T_c$&$1.3T_c$&$2T_c$\\ 
\end{tabular}
\end{ruledtabular}
\end{minipage}
\caption{Upper bound on dissociation temperatures.}
\end{center}
\label{tab:diss}
\end{table}

In conclusion, we determined quarkonia spectral functions 
in the quark-gluon plasma using a potential model with two choices for the potential, both motivated by
lattice QCD results on the free energy of a static quark anti-quark pair. We found that, due to color screening, 
for the first chosen potential most quarkonia states, except the $\Upsilon$, dissolve at temperatures close
to that of deconfinement. 
For the most extreme
potential which is still compatible with lattice data, resonance structure in 
the spectral functions exists up to higher temperatures. This potential provides an upper limit on the binding energy.
Using the binding energy we calculate the width of various states, and
give upper bounds on their dissociation temperatures which are significantly lower
than previous estimates. As such, the model proposed in \cite{karsch06}, where $J/\psi$ suppression 
is due only to melting of the $\chi_c$ and $\psi'$ states, cannot  explain the nuclear
modification factor $R_{AA}$ measured in the experiments since color screening dissolves the $J/\psi$. 
On the other hand, the enhancement of the spectral function near the threshold shows that the heavy quarks and antiquarks remain strongly correlated in the plasma even though they do not form a bound state. This correlation could
lead to the regeneration of some quarkonium states when the plasma converts to hadronic matter increasing $R_{AA}$ values above expectations from screening alone. The quark and antiquark may even reform into a higher excited state. For a quantitative description of $R_{AA}$, a model calculation of regeneration effects is needed. More precise calculations of the spectral function and detailed lattice calculations of the static quark-antiquark correlators will also be helpful. 

{\bf Acknowledgments:}
We are grateful to J. Casalderrey-Solana for his contribution at the early stages of this work.
We thank D. Kharzeev, L. McLerran and P. Sorensen for careful reading of the manuscript and valuable comments. 
This work has been supported by U.S. Department of Energy under Contract No. DE-AC02-98CH10886.


\end{document}